\documentstyle[11pt,newpasp,twoside,epsf]{article}
\def\edcomment#1{\iffalse\marginpar{\raggedright\sl#1\/}\else\relax\fi}
\reversemarginpar

\begin{document}
\newcommand{\cthead}[1]{\multicolumn{1}{c}{#1}}
\newcommand{\ks}{km~s$^{-1}$}
\newcommand{\kss}{km~s$^{-1}$ }
\newcommand{\etal}{et~al.}

\title{Space-VLBI observations of OH masers}
 \author{V.I. Slysh, M.A. Voronkov, I.E. Val'tts}
\affil{Astro Space Center, Profsouznaya st. 84/32, 117810 Moscow, Russia}
\author{V. Migenes,}
\affil{University of Guanajuato, Department of Astronomy, Apdo
Postal 144, Guanajuato, CP36000, GTO, Mexico}
\author{K.M. Shibata, T. Umemoto,}
\affil{National Astronomical Observatory, 2-21-1 Osawa, Mitaka, Tokyo 181,
Japan}

\begin{abstract}
We report on the first space-VLBI observations of the OH
masers in two main-line OH transitions at 1665 and 1667 MHz. The
observations involved the space radio telescope on board the
Japanese satellite {\it HALCA} and an array of ground radio telescopes.
The maps of the maser region and images of individual maser spots were
produced
with an angular resolution of 1 mas, which is several times higher
than
the angular resolution available on the ground. The maser spots were
only partly resolved and a lower limit to the brightness temperature
$6\times10^{12}$~K was obtained. The masers seem to be located in the
direction of low interstellar scattering.
\end{abstract}

\section{Introduction}
OH masers were the first to be discovered as a new class of astronomical
phenomenona exhibiting intense, narrow-band, polarized, variable
emission in interstellar molecular transitions. The source
of this emission was shown to be small, compact clumps of neutral gas, at
the periphery of compact \hbox{H\,{\sc ii}} regions created by newly born
massive O stars. Another class of OH masers is related to evolved low-
and medium-mass stars, and is not considered here. High angular resolution
study of OH masers showed that they consist of a number of bright compact
maser spots separated by several arcseconds. The spots themselves are from
2 to 70 mas in extent, and in some masers are barely resolved,
even with the highest angular resolution available from ground-based
VLBI arrays.

Current models of OH maser emission deal with gas condensations of
particle density 10$^7$~cm$^{-3}$, kinetic temperature 100~K,
illuminated by the far-infrared emission from hot dust. With Doppler
velocity gradient FIR-line overlap produces population inversion
responsible for the main-line emission at 1665~MHz and 1667~MHz.
A magnetic field of about 10~milligauss is sufficient to produce
Zeeman splitting of lines and strong elliptical polarization.

The goal of the high angular resolution study
of OH masers is to determine relative position of the spots, to measure the
intensity, and to map the
spots. Determination of the size and the shape of the maser spots would
be very helpful for understanding the maser emission mechanism and its
limiting brightness temperature. The shape of the maser spots may be
indicative of the type of physical phenomena responsible for the origin
of the maser emission.

Several possible sites for the origin
of maser emission - 
such as the shocks at the border of \hbox{H\,{\sc ii}}
regions, or the interaction region between a molecular outflow and
ambient molecular clouds, as well as protoplanetary accretion disks around
young star -- have been suggested.
The maser spot shape and size are important properties of the maser
emission, provided they are
intrinsic to the source, and not caused by a propagation effect such as
interstellar scattering.

There is a widespread opinion that OH masers
are heavily scattered by interstellar turbulence, and that observed
images of maser spots are scattering-broadened images of essentially
point-like sources. This view is based on data obtained from the study
of scattering of pulsars and continuum sources. The importance of
scattering for OH masers is due to their location within the thin
galactic disk where the scattering is most severe.

Slysh et al. (1996) used a three-station VLBI network to measure the
angular size of three compact OH masers. With a fringe separation of
4.2~mas and high signal-to-noise ratio data, the angular size of several
spectral features in the two main-line OH transitions was measured
simultaneously. The
measured angular size was in the range from 1.4$\pm$0.4~mas to
4.3$\pm$0.1~mas. It was found that the angular size of the spectral
features with the same radial velocity, presumably originating in the same
region, was larger in the 1667-MHz line than in the 1665-MHz line. This cannot
be the result of scattering, since the scattering size should be virtually the 
same at these two close frequencies.
It was concluded that the measured
angular size at 1667~MHz, 4.3$\pm$0.1~mas for OH34.26$+$0.15 and
3.5$\pm$0.2~mas for W48, is intrinsic to the sources. These two masers
together with OH45.47+0.13 exhibit an angular size for
spectral features which is much less than the average broadening estimated
from pulsar measurements. The conclusion of that study was that the
distribution of the scattering material in the Galaxy is patchy, and that
these three masers are located in a direction of low interstellar
scattering. Still higher angular resolution was achived in space-VLBI
observations of OH masers using a 8-m telescope on board of the
Japanese spacecraft {\it HALCA} and an array of large ground telescopes
(Slysh et al.~2001). The observations of
OH34.26$+$0.15 with the synthesized beam of about 1~mas partially resolved
maser spots and confirmed the low interstellar scattering in the direction of
this maser. In this communication we present additional space-VLBI data
for OH masers OH34.26$+$0.15, W48 and CepA. 

\section{Observations and data reduction}

Three OH masers were observed as a part of the
Key Science Program of the
Japanese Satellite {\it HALCA} in 1998, by the space-ground
very long baseline interferometer.
The satellite radio
telescope has an 8-m diameter deployable parabolic mirror and uncooled L-band
receiver, orbiting the Earth with a 6-h period, apogee 21000~km and
perigee 560~km~(Hirabayashi et al. 1998).
The ground radio telescope array 
 for OH34.26+0.15 was: the 70-m DSN telescope Tidbinbilla in Australia,
the phased 6$\times$22-m Australia Telescope, the 22-m Mopra Telescope in
Australia,
the Shanghai 25-m Telescope in China,
the 64-m Telescope in Usuda, Japan,
and the 
64-m Telescope in Bear Lakes near Moscow. It was the first time
that the Bear Lakes Telescope participated in space-VLBI observations;
for W48: Australia Telescope, Mopra, Shanghai, 26-m telescope in South
Africa (Hartebeesthoek), 26-m telescope in Hobart (Tasmania), 32-m
telescope in Noto (Italy);
for CepA: 70-m telescope in Robledo (Spain) and 10x25-m VLBA (USA).

Left circular polarization data over a 16-MHz
bandwidth centered on the 1665-MHz and 1667-MHz OH main lines
were recorded
using S2 recorders at each station. The data were correlated using the
Canadian S2 space-VLBI correlator. The 16 MHz data were filtered using a 
bandpass filter and then resampled before correlation in order to
attain a spectral resolution of 488 Hz~channel$^{-1}$\-  (0.088~\ks)
over a bandpass of 500 kHz.
The CepA data were recorded on VLBA recorders and
correlated on the VLBA correlator in Socorro, USA,
with the 16-MHz band divided in 1024 spectral channels, giving a
spectral resolution of 15.625~kHz per channel (2.81~\kss).
Amplitude calibration was achieved using the 'gain' and
T$_{sys}$ data obtained from ground observatories
and from the VSOP operations group. The post-correlation data reduction
and imaging was
performed at the Astro Space Center using the {\sc aips} package of NRAO.
The strongest and most compact
spectral feature was taken as a reference in the self-calibration
process. Its position relative to the correlating position
was determined by the fringe rate
method, using the ground array.

\section{Results}

\subsection{OH34.26+0.15}
\begin{figure}[!hbt]
\plottwo{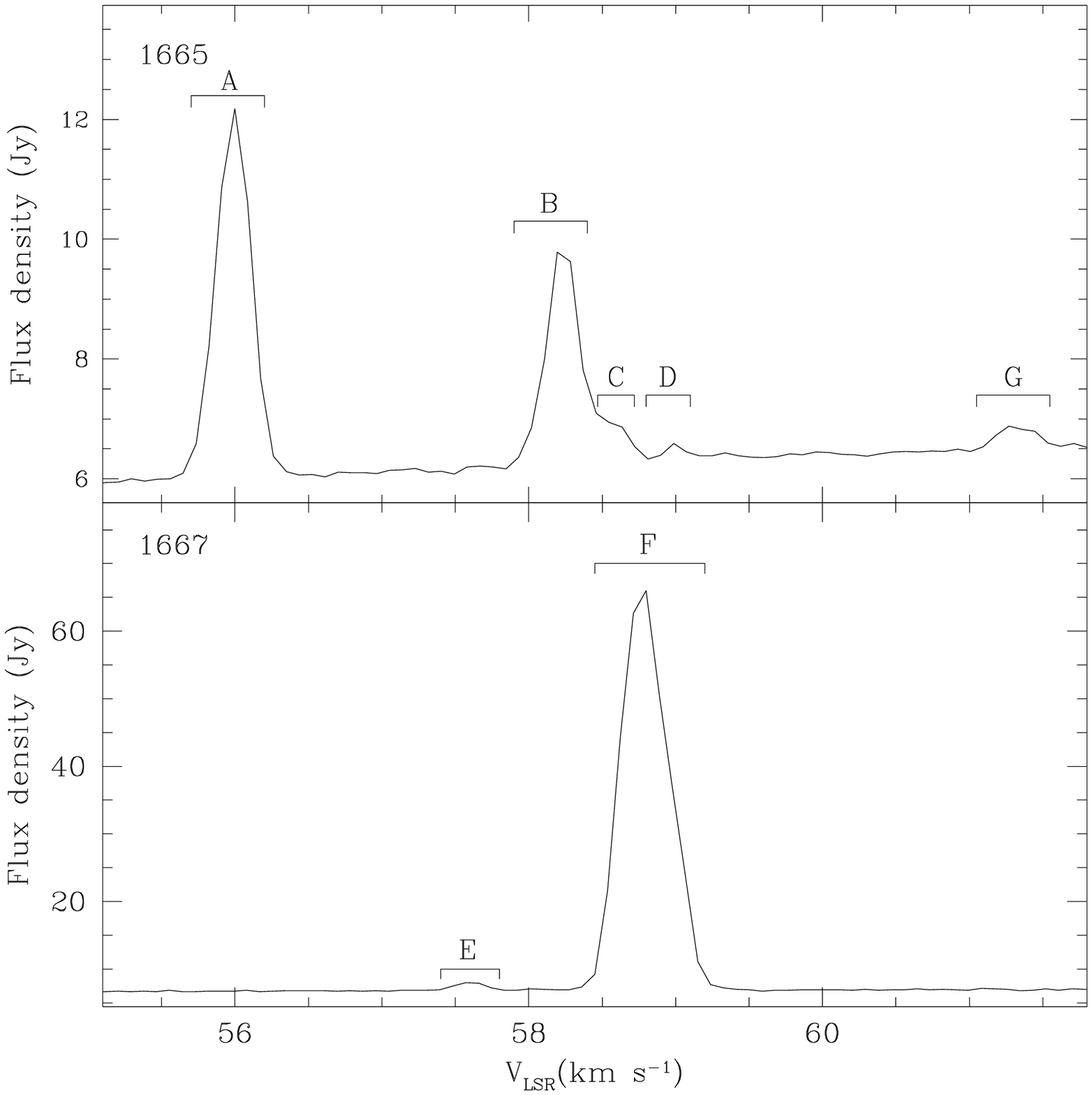}{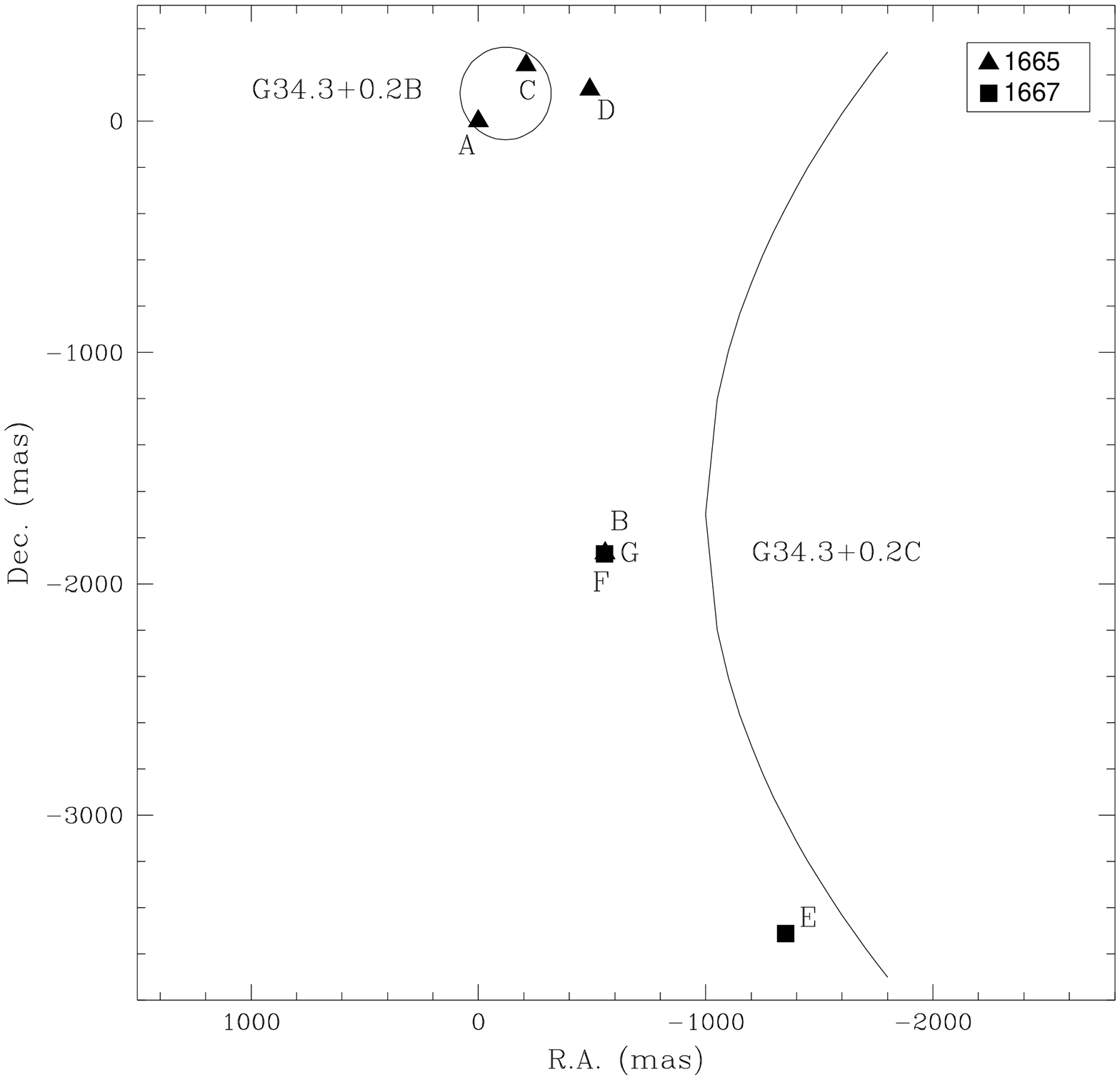}
\caption{(left) Correlated spectrum of OH34.26$+$0.15 on the short
base line AT -- Mopra. Upper panel: the 1665-MHz line; lower panel:
the 1667-MHz line. (right) Map of OH34.26$+$0.15. Triangles are
1665-MHz features, squares are 1667-MHz features. A circle is ultracompact
H{\sc II}  region G34.3$+$0.2B and a line shows position of the 
cometary H{\sc II} region G34.3$+$0.2C.}
\end{figure}

The correlated LCP spectra of OH34.26+0.15 on the short Australia
Telescope -- Mopra baseline on which the source is unresolved,
in both main-line OH
transitions, is shown in Fig.~1 (left). Five spectral features A--D
and G
are present in the 1665-MHz spectrum, and two features E and F in the 1667~MHz
spectrum; their relative position is shown in Fig.~1 (right).
All components are partly resolved with the beam of
the space-ground or ground
only array, but most of the components may have structure and consist of
several smaller components.
Feature B consists of four separate
closely spaced spots.
Feature F (1667~MHz) coincides with feature B within
the measurement errors, and its shape is similar to the shape of
feature B.
These two features may come from the same maser condensation, a
supposition that can be
tested with higher accuracy relative 1665/1667 position measurements.
If this is the case the radial velocities of features B and F must coincide,
and the
observed velocity difference of 0.5~\kss can be attributed to a difference
between Zeeman splittings of 1665-MHz and 1667-MHz lines because of
different g-factors of the two transitions. If the observed features B and F are
$\sigma$-components of Zeeman pairs (both are left circular polarized), one
can calculate that the velocity difference of 0.5~\kss corresponds to
the magnetic field of B=4.2~milligauss, which is a value commonly found in
OH masers.

OH34.26$+$0.15 is located at a distance of 3.8~kpc
near a cometary
\hbox{H\,{\sc ii}} region and near two ultracompact {H\,{\sc ii}} regions.
Maser spots A, C and D coincide with the northern ultracompact
\hbox{H\,{\sc ii}} region G34.3$+$0.2B, while spots B, E, F and G
coincide with the cometary \hbox{H\,{\sc ii}} region
G34.3+0.2C. Our map of OH
components (Fig.~1) is consistent with the VLA map of the
same source obtained by Gaume \& Mutel~(1987) in 1985.

The most compact component is feature A, which
is known to be left hand circularly polarized.
It was
resolved in one direction; the angular size in this direction is
2.0~mas and less than 0.5~mas in the perpendicular direction.
Feature A was fitted with an ellipse  in position angle 70\deg$\,$ with
the ratio of major to minor axes greater than 4. This size is a result of the
fitting and deconvolution with the beam of the image. 
As an almost independent check of the size of
feature A, we made a direct fit of a Gaussian model to {\it uv} data using only
fringe amplitude information. This procedure is independent in the sense
that it does not use the self-calibration and global fringe fitting.
The best fit to fringe amplitudes from three telescopes AT, Tid and
{\it HALCA} was achieved with an elliptical Gaussian, having major and minor
axes 2.6$\pm$0.5 and 0.3$\pm$0.3~mas and a position angle 61\fdg5$\pm$3\fdg2
which, within the errors, is the same as obtained
from the image. From our data it is not possible to say whether maser
spot A has a stripe-like shape, or is composed of a chain of
point-like sources. To determine this, one needs higher angular resolution
and higher signal-to-noise data in order to obtain maps with a higher
dynamic range.

The upper limit of the angular size of feature A corresponds
to the lower limit of its brightness temperature of 6$\times10^{12}$~K. Other
features have somewhat larger sizes and lower brightness temperature limits,
although it cannot be excluded that they consist of several unresolved
subcomponents.

\subsection{W48}
\begin{figure}[tb]
\plottwo{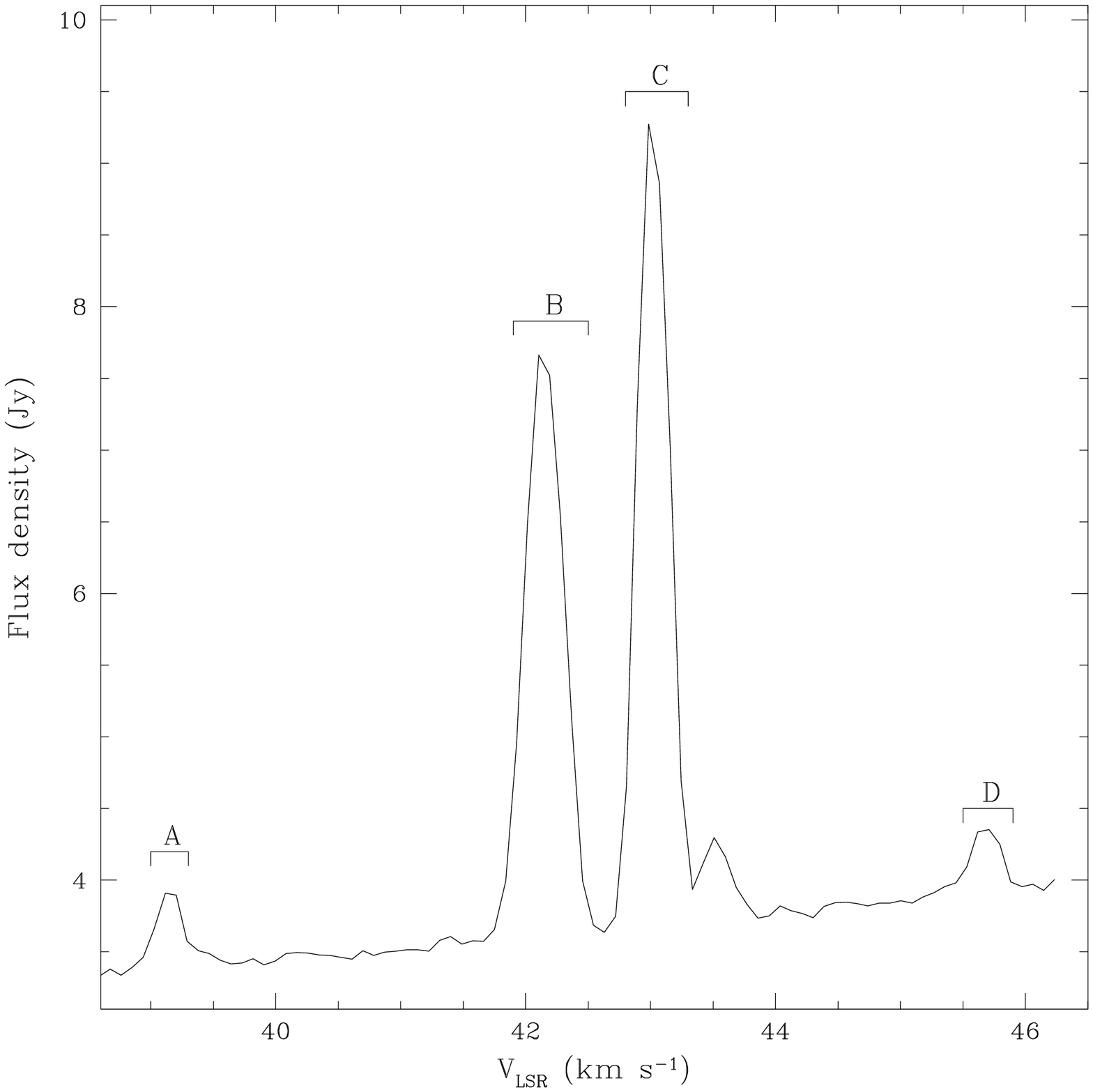}{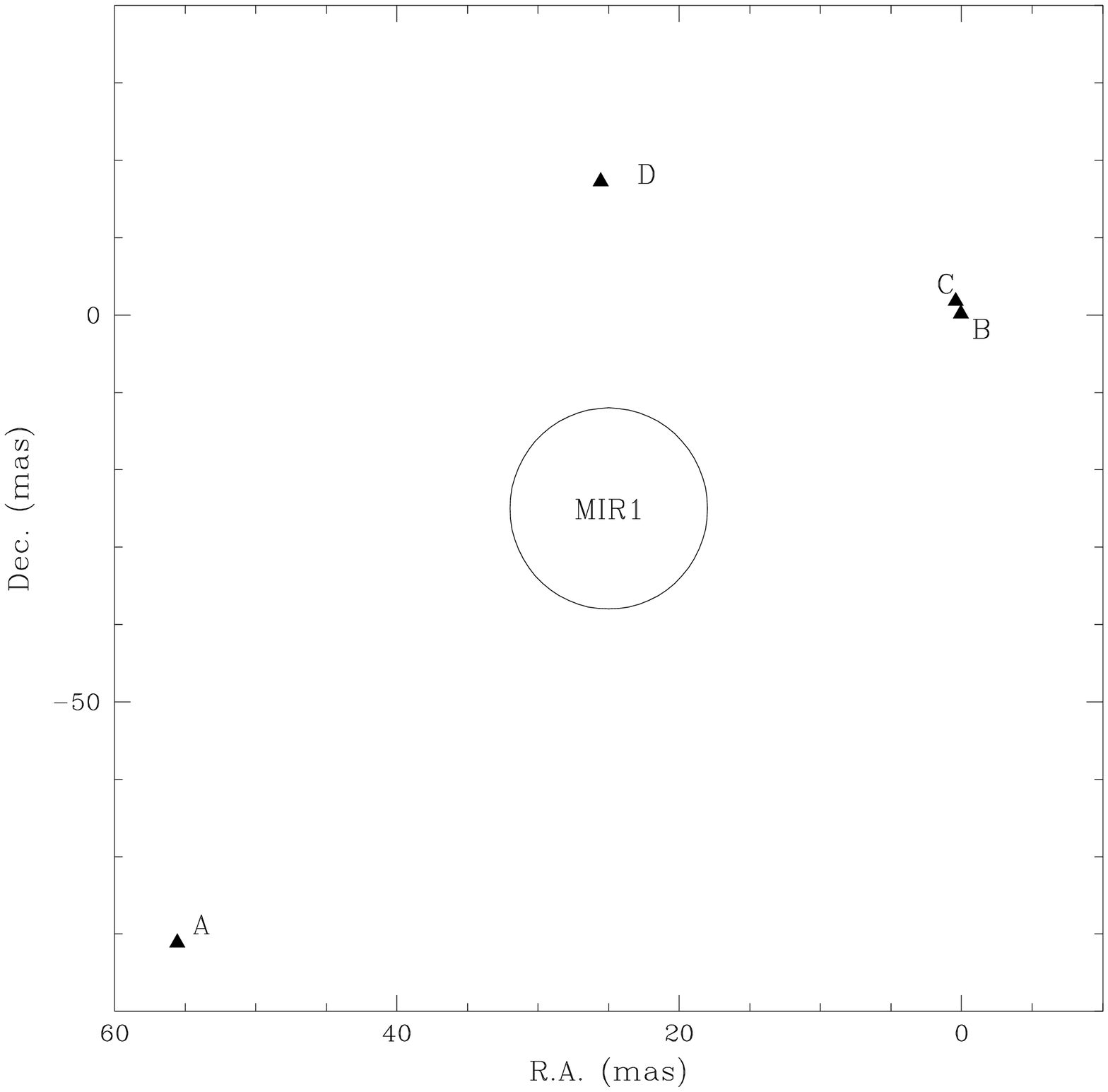}
\caption{(left) Correlated spectrum of W48 at 1665~MHz on the short
base line AT -- Mopra. (right) Component map. 
Circle at the image center represents a mid-infrared source
associated with this maser.}
\end{figure}

The correlated 1665-MHz LCP spectrum on the short baseline
is shown on Fig~2~(left). There are four spectral feature in the spectrum,
and on the map four spots correspond to these features (Fig.~2~(right)).
The largest separation between the spots is 100~mas, or 310~AU at
the distance 3.1~kpc. In 1667-MHz line there is a single spectral
feature at LSR velocity 42.2~\kss, the position of the 1667-MHz
feature has not been measured. The 1667-MHz feature is resolved, with
angular size 13.0x2.3~mas at the position angle 115$^\circ$. The minor
axis in fact is an upper limit, therefore the 1667-MHz spot may be a
13~mas long filament of the width less than 2.3~mas, or a chain of
closely spaced point sources. All 1665-MHz spots are unresolved with
the VSOP-to-ground baselines, with the syntheesized beam 2.1x0.8~mas.
The upper limit of the angular size of the brightest feature B
is 1~mas, which corresponds to the linear size 3~AU.

There is no ultracompact region at the position of the OH maser,
the nearest ultracompact region G35.20$-$1.74 is 20$''$ away.
However, there is a mid-infrared source MIR~1 (Persi et al. 1997)
coinciding also with H$_2$O maser HC1 (Hoffner \& Churchwell 1996).
Its luminosity is
240-300~L$\odot$ and can be due to a pre-main sequence star or a
young star of the mass of about 4~M$\odot$. The observed
velocity spread of the  OH maser 6.6~\kss and the separation
of 310~AU are consistent with OH masers being gravitationally
bound by the 4~M$\odot$ star in the middle. The lower limit of
the brightness temperature of the brightest feature B is 4.2x10$^{12}$~K.

\subsection{CepA}
\begin{figure}[hbt]
\plottwo{slysh1fig3a.ps}{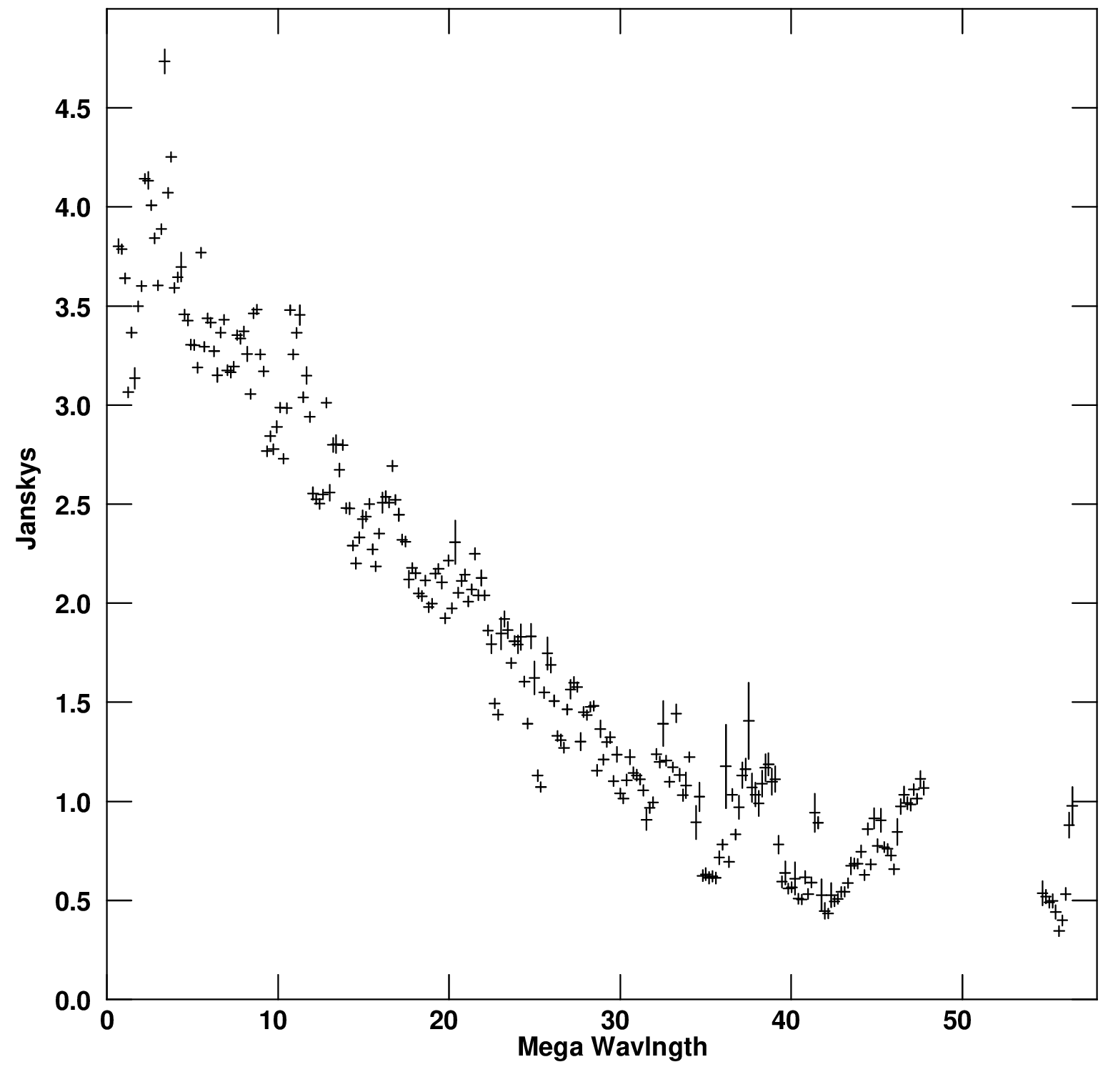}
\caption{(left) Component map of CepA. Position and size of ultracompact
H{\sc II} regions is shown by ellipses. (right) Visibility plot
versus baseline length for the feature Q (V$_{LSR}=-$13.9~\kss).}
\end{figure}

CepA was observed with {\it HALCA} and ground array consisting of the VLBA
and 70-m telescope in Robledo, Spain. The tape recorder was VLBA recorder,
and correlation has been done at the VLBA correlator, with 16~MHz band
divided into 1024 spectral channels. This gave a very poor spectral
resolution of 2.81~\kss, which was inadequate to resolve separate spectral
features. Also, their amplitude was reduced due to the spectral smearing
of narrow spectral features (typically 0.3~\kss) in the wide spectral
channels. Nevertheless, one can distinguish between several maser spots
with VLBI mapping even if they are in the same spectral channels. Both
1665-MHz and 1667-MHz lines were in the same band which allowed to image
both lines on the single map. The spot map of CepA is shown on Fig.~3~(left).
22 spots (4 spots at 1667~MHz and 18 spots at 1665~MHz) have been
mapped with the ground array which provided the synthesized beam
3.3x4.0~mas. Only one 1667-MHz spot~W coincides with the 1665-MHz spot~O,
within 1~mas. 
Other spots as all 1665-MHz spots are scattered
over an area of about 2$''$.5 by 3$''$.5, or 1800 by 2500~AU
at the distance of 0.7~kpc. This map is in a good agreement with the
MERLIN map obtained by Migenes et al. (1992).
In the same area there are several ultracompact HII regions (HW2, HW3c
and D) shown on Fig.~3~(left) as ellipses (Torrelles et al. 1998).
They are excited by massive B0-B1 stars each, of the mass 12-13M~$\odot$.
The maser spots can be gravitationally bound to these stars, either to
a particular star (spots S,R,Q,W,O to HW3d, spots K,J,E,F,G,H,I to HW2),
or to the common gravitational potential of all the stars (the rest of
the spots). The stars themselves can be gravitationally bound forming a
multiple star system. At the mean distance of 1000~AU the orbital velocity
for 13~M$\odot$ star is 3.5~\kss which is about the same as the velocity
spread of OH maser features. The maser spots have been completely resolved
on all ground-to-space baselines. Fig~3~(right) shows correlated flux versus
baseline  dependence for the brightest spot~Q. The decline of the
correlated flux with the baseline corresponds to a Gaussian spot size
of 3.8~mas. The image of this spot (not shown) corresponds to a Gaussian
source 5.1x3.4~mas. The size of the rest of 1665-MHz spots is of the
same order. The 1667-MHz spots are much larger: the major axis is 18-20~mas,
and the minor axis is from 4 to 13~mas. The linear size of the spots
is from 3 to 14~AU, and the brightness temperature of the spot~Q
is 3.5x10$^{12}$~K.

\section{Summary and conclusions}

OH masers are associated with single massive stars or multiple star systems.
The lowest mass star is 4~M$\odot$ star in W48, and the highest mass
25~M$\odot$ is in OH34.26$+$0.15. The maser spots are gravitationally
bound to the stars, and may be on Keplerian orbits in single star systems,
or on  complicated orbits in multiple star systems, at the distance 150~AU
to 3000~AU. The maser spots are partly resolved by the space-to-ground
interferometer. The most compact are maser spots in W48. In CepA which is
the nearest OH maser the maser spots were completely resolved. The linear
size of the maser spots is several astronomical units in all masers.
1665-MHz masers are a factor of 3 more compact than 1667-MHz masers.
The brightness temperature ranges from 3.5x10$^{12}$~K in CepA to more
than 6x10$^{12}$~K in OH34.26$+$0.15. The interstellar scattering in the
direction of these masers does not increase observed angular size and is
less than 1~mas.

\section{Acknowledgements}

We gratefully acknowledge the VSOP project, which is led by the Japanese
Institute of Space and Astronomical Science in cooperation with many
organizations and radio telescopes around the world. The National Radio
Astronomy Observatory is a facility of the National Science Foundation,
operated under a cooperative agreement by Associated Universities, Inc.
VIS, MAV and IEV acknowledge support from the INTAS (grant No. 97-11451)
and from the Russian Foundation for Basic Research (grant No. 01-02-16902).

\end{document}